\documentclass{PoS}
\usepackage{graphicx,amsmath,alltt}

\DeclareSymbolFont{ttoperators}{OT1}{cmtt}{m}{n}
\newcommand\xCode[1]{{%
  \mathcode`\"="0\the\symttoperators22%
  \mathchardef\$="4\the\symttoperators24%
  \mathcode`\(="4\the\symttoperators28%
  \mathcode`\)="5\the\symttoperators29%
  \mathcode`\/="0\the\symttoperators2F%
  \mathcode`\[="4\the\symttoperators5B%
  \mathcode`\]="5\the\symttoperators5D%
  \mathchardef\{="4\the\symttoperators7B%
  \mathchardef\}="5\the\symttoperators7D%
  \ensuremath{\mathtt{#1}}}}
\newcommand\Code[1]{\texttt{#1}}
\newcommand\eg{e.g.\ }
\newcommand\ie{i.e.\ }
\newcommand{\rd}{\mathrm{d}}
\newcommand\greyed[1]{\special{ps: .5 setgray}#1\special{ps: 0 setgray}}
\newcommand{\ket}[1]{\left| #1\right\rangle}
\newcommand{\bra}[1]{\left\langle #1\right|}

\title{FormCalc 7.5}
\ShortTitle{FormCalc 7.5}

\author{S.~Agrawal \\
LNM IIT, Rupa Ki Nagal, Post-Sumel, Via Jamdoli, Jaipur--302031, India \\
E-mail: shivam@mpp.mpg.de}

\author{\speaker{T.~Hahn} \\
MPI f\"ur Physik, F\"ohringer Ring 6, D--80805 Munich, Germany \\
E-mail: hahn@mpp.mpg.de}

\author{E.~Mirabella \\
MPI f\"ur Physik, F\"ohringer Ring 6, D--80805 Munich, Germany \\
E-mail: mirabell@mpp.mpg.de}

\abstract{We present additions and improvements in Version 7.5 of 
FormCalc, most notably OPP methods, Output in C, MSSM initialization via 
FeynHiggs, and Analytic tensor reduction, as well as a parallelized Cuba 
library for numerical integration.
\\ \hbox to\hsize{\hfill Report MPP-2012-136}
}

\FullConference{Loops and Legs in Quantum Field Theory -- 11th DESY 
Workshop on Elementary Particle Physics,\\
April 15--20, 2012 \\
Wernigerode, Germany}

\begin{document}

\section{Introduction}

The Mathematica package FormCalc \cite{FormCalc} simplifies Feynman 
diagrams up to one-loop order generated by FeynArts \cite{FeynArts}. 
It provides both the analytical results and can generate code for the 
numerical evaluation of the squared matrix element.

Mathematica's powerful language enables users to easily inspect and 
modify results and should be considered a feature, not the deficiency 
other packages claim vindicates their use of \eg Python.

This note presents improvements and additions in FormCalc 7.5 and the 
numerical integration package Cuba 3 \cite{Cuba}, which is also included
in FormCalc:
\begin{itemize}
\item Unitarity methods (OPP),

\item Parallelization of helicity loop,

\item Output in C and Improved code generation,

\item Command-line parameters for model initialization,

\item MSSM initialization via FeynHiggs,

\item Analytic tensor reduction,

\item Auxiliary functions for operator matching,

\item Built-in parallelization in Cuba.
\end{itemize}

\section{Unitarity methods (OPP)}

FormCalc 7 can generate code which uses the OPP (Ossola, Papadopoulos,
Pittau \cite{OPP}) unitarity methods as implemented in the two libraries
CutTools \cite{CutTools} and Samurai \cite{Samurai}.  Rather than
introducing Passarino--Veltman (PV \cite{PaVe}) tensor coefficient
functions, the entire numerator is placed in a subroutine, as in:
$$
\varepsilon_1^\mu\varepsilon_2^\nu B_{\mu\nu}(p, m_1^2, m_2^2) =
  B_{\mathrm{cut}}(2, N, p, m_1^2, m_2^2)\,,
\qquad\text{where}\qquad
N(q_\mu) = (\varepsilon_1\cdot q) \: (\varepsilon_2\cdot q)\,.
$$
The numerator subroutine $N$ will be sampled by the OPP function 
($B_{\mathrm{cut}}$ in this example).  The first argument of
$B_{\mathrm{cut}}$, 2, gives the maximum power of the integration 
momentum $q$ in $N$.

Subexpressions of the numerator function (coefficients, summands, etc.) 
which do not depend on $q$ are pulled out and computed once, ahead of 
invoking the OPP function, using FormCalc's abbreviationing machinery 
\cite{FCabbr}.  In particular in BSM theories, these coefficients can be 
lengthy such that pulling them out significantly increases performance.

The CutTools and Samurai libraries have minor differences in calling 
conventions but are otherwise similar enough to let the preprocessor 
handle the switching.  That is, one does not need to re-generate the 
Fortran code in order to link with the other library.  Specifically, the 
following steps must be taken in order to use the OPP method in 
FormCalc:
\begin{itemize}
\item The amplitudes must be prepared with \Code{CalcFeynAmp[..., OPP 
$\to n$]} ($n < 100$).

\item In the generated code, the OPP library (CutTools or Samurai) must 
be chosen and the declarations in \Code{opp.h} be included.  This is 
most conveniently done in \Code{user.h}, in the following structure:
\vspace*{-3.2ex}
\begin{alltt}
\greyed{ }
\greyed{#ifndef USER_H}
\greyed{#define USER_H}
\greyed{* declarations for the whole file (e.g. preprocessor defs)}
#define SAMURAI                \greyed{\textrm{\emph{(or \texttt{CUTTOOLS})}}}
\greyed{#else}
\greyed{* declarations for every subroutine}
#include "opp.h"               \greyed{\textrm{\emph{(necessary for OPP)}}}
\greyed{#include "model_sm.h"}
\greyed{#endif}
\end{alltt}
\end{itemize}
We have presently compared a handful of $2\to 2$ and $2\to 3$ scattering 
reactions, both QCD and electroweak, and found agreement to about 10 
digits between PV and OPP, with CutTools and Samurai delivering results 
of similar quality.  This shows that the method is working.

The performance is lagging quite a bit, however, at least when applying 
the OPP method naively and for lower-leg multiplicities.  The principal 
difference is that the numerator function imposes a helicity dependence 
on the OPP function such that, unlike the PV tensor coefficients, it 
cannot be hoisted out of the helicity loop.

The following improvements have been made to optimize performance:
\begin{itemize}

\item Our implementation admits mixing PV decomposition with OPP in the 
sense that one chooses an integer $n$ starting from which an $n$-point 
function is treated with OPP methods.  For example, \Code{OPP $\to$ 4} 
means that $A$, $B$, $C$ functions are treated with PV and $D$ and up 
with OPP.  A negative $n$ indicates that the rational terms for the OPP 
integrals shall be added analytically whereas else their computation is 
left to the OPP package.

\item The number of OPP calls turns out to be more detrimental to 
performance than the complexity of the numerators.  The simplification 
strategy for OPP integrals is thus to join, rather than split (as for 
PV), denominators.  A loop integral whose denominators form a complete 
subset of another are joined with the latter, as in
$$
\frac{N_4}{D_0 D_1 D_2 D_3} + \frac{N_3}{D_0 D_1 D_2}
\to \frac{N_4 + D_3 N_3}{D_0 D_1 D_2 D_3}
$$
Furthermore, simplifications that break up loop integrals are suppressed, 
such as the cancellation of $q^2$-terms, \eg
$$
\frac{q^2}{(q^2 - m^2) D_1 D_2} \not\to
\frac 1{D_1 D_2} + \frac{m^2}{(q^2 - m^2) D_1 D_2}\,.
$$

\item Profiling the code pointed us to inefficiencies in the evaluation 
of fermion chains.  In older FormCalc versions, the computation 
proceeded through nested invocations of elementary operations 
(2-component matrix--vector and vector--vector products \cite{FCopt}).  
Inlining these functions in a portable way in Fortran was syntactically 
not straightforward, so we switched to single function calls for an 
entire chain, as in:
\begin{align*}
\bra{u}\sigma_\mu\overline{\sigma}_\nu&\sigma_\rho\ket{v}
k_1^\mu k_2^\nu k_3^\rho
=: \bra{u} k_1 \overline{k}_2 k_3\ket{v}
\\
\text{old} &= \Code{SxS(}u,\,
  \Code{VxS(}k_1,\, \Code{BxS(}k_2,\, \Code{VxS(}k_3,\,
  v \Code{))))} \\
\text{new} &= \Code{ChainV3(}
   u, k_1, k_2, k_3, v
   \Code{)}
\end{align*}
As the profiler no longer `sees' these now-inlined functions, we cannot 
quote a concrete figure for the performance gain; a naive before--after
comparison of wall-clock time indicates an improvement on the order of 
5--10\%, however.

\item To avoid the evaluation of integrals whose prefactor is known to 
be exactly zero from helicity considerations we added an ``helicity 
delta'' argument to each OPP integral, for example:
$$
\Code{Dcut(1\,-\,Hel1, rank, num, \dots)},
$$
which will not be evaluated if \Code{Hel1} is 1.

\end{itemize}

%%%%%%%%%%%%%%%%%%%%%%%%%%%%%%%%%%%%%%%%%%%%%%%%%%%%%%%%%%%%%%%%%%%%%%

\section{Parallelization of Helicity Loop}

Perhaps the most obvious way to address the OPP slowdown is to 
parallelize the loop over the helicities.  Code generated by FormCalc is 
in fact well suited for this as FormCalc does not insert explicit 
helicity states in the algebra already \cite{FCopt}.  That is, the 
amplitude is a numerical function of the helicities $\lambda_i$ and not 
a bunch of (different) functions for each helicity combination,
$$
\mathcal{M} = \mathcal{M}(\lambda_1, \lambda_2, \dots)
\neq \{\mathcal{M}_{--\cdots},\ \mathcal{M}_{+-\cdots},\ 
     \mathcal{M}_{-+\cdots},\ \mathcal{M}_{++\cdots}\}
$$
In computer science this is known as a Single Instruction Multiple Data 
(SIMD) design since a single code ($\mathcal{M}$) is independently 
run for multiple data ($\lambda_i$), and is conceptually easy to 
parallelize.

Our process model has one master and $N$ workers on an $N$-core system.  
The master is in charge of the non-helicity-dependent parts of the 
computation and coordinates the workers, \ie starts/stops them and 
distributes/collects the data.  Currently we use the same 
\Code{fork}/\Code{wait} technology as in Cuba, with socketpair I/O for 
the data transmission (see Sect.~\ref{sect:cuba}).

The implementation is brand new and speed-up measurements are not yet 
available.  Unless the helicity loop accounts for a significant part of 
the computation time, however, the achievable overall speed-ups may well 
be limited as the master performs the non-helicity-dependent work 
single-threaded.  Thus, we expect OPP to gain more from parallelization 
than the PV method.

Unless one is evaluating a single phase-space point only, the helicity 
parallelization clearly competes for compute cores with Cuba, and it is 
at present an open question which is the optimal strategy for assigning 
the cores.  Then again, the current \Code{fork}/\Code{wait} code can be 
regarded as a proof-of-concept implementation and may be substituted \eg 
by a GPU version in the future, for which most of the organizational 
groundwork is then already laid.  For instance, a sizable part of the 
present effort went into grouping the variables into helicity-dependent 
and -independent ones, to minimize communication overhead between the 
master and the workers.  Likewise, new versions of the LoopTools 
functions had to be introduced to obtain control over the cached loop 
integrals.

The parallelization is enabled in the code by
\begin{verbatim}
   #define PARALLEL
\end{verbatim}
which is usually placed in \Code{user.h}.  The actual number of cores 
used can be specified in the environment variable \Code{FCCORES}.  If 
\Code{FCCORES} is not set, all free cores (total cores minus current 
system load) are used.  Note that, at least for now, FormCalc is not 
aware of the number of cores taken by Cuba, \eg the \Code{CUBACORES} 
variable.

%%%%%%%%%%%%%%%%%%%%%%%%%%%%%%%%%%%%%%%%%%%%%%%%%%%%%%%%%%%%%%%%%%%%%%

\section{C Output and Improved code generation}

Code generation in C99 is available in FormCalc 7.5 next to the 
traditional Fortran output.  This feature is enabled with
\begin{verbatim}
   SetLanguage["C"]
\end{verbatim}
The generated code is binary-compatible with Fortran, \ie its object 
files can be linked directly to a Fortran program.  The C code 
accordingly observes Fortran calling conventions (pointers only) and
uses underscore-suffixed lowercase function and struct names.  One 
temporary setback is that there is no automatic translation yet of the 
declarations part of the FormCalc driver modules, \eg of the model 
parameters, which are obviously necessary for compiling the C code.

The advantages of C code are threefold:
\begin{itemize}
\item It makes integration of generated code into existing C/C++ 
packages easier (no linking hassles).

\item It simplifies GPU programming; for Fortran, only a single, 
commercial, compiler (PGI) currently targets the GPU.

\item One can take advantage of C's \Code{long double} data type which, 
at least on Intel x86 hardware, gives an additional 2--3 digits of 
precision at essentially no extra cost.  Extended real data types in 
Fortran, if available, are often IEEE-754-compliant \Code{REAL*16} 
emulated in software.  Only gfortran 4.6+ offers the \Code{REAL*10} 
Fortran equivalent of \Code{long double}.
\end{itemize}

Improvements have been made to the Fortran code generation as well:
\begin{itemize}
\item Loops and tests are handled through preprocessor macros, \eg
\begin{verbatim}
   LOOP(var, 1,10,1)
   ...
   ENDLOOP(var)
\end{verbatim}
This aids automated substitution with \eg \Code{sed}.  It also enhances
readability (\Code{ENDLOOP} with a variable name rather than an 
incognito \Code{enddo}) and makes the `look' of the C and Fortran code 
fairly similar.

\item Likewise, the main subroutine \Code{SquaredME.F} is now sectioned 
by comments.  For example, the variable declarations are enclosed in
\begin{verbatim}
   * BEGIN VARDECL
   ...
   * END VARDECL
\end{verbatim}

\item The generated code and the driver files are consistently 
formulated in terms of the newly introduced \Code{RealType} and 
\Code{ComplexType} data types, by default equivalent to \Code{double 
precision} and \Code{double complex}, respectively.  Note that 
capitalization matters as these words are substituted by the 
preprocessor.  This introduces a level of abstraction which makes it 
easier to \eg work with a different precision.
\end{itemize}

\section{Command-line parameters for model initialization}

FormCalc includes a suite of so-called driver programs to manage the 
automatically generated code for computing the squared matrix element. 
They parse the command line, initialize model constants, set up phase 
space, etc.

In particular the driver modules for the initialization of the model 
parameters and luminosity calculation (which includes \eg the setup of 
PDFs used in hadronic reactions) had no access to the command-line 
arguments so far and could use only variables supplied by the user in 
the main control program \Code{run.F}.  In other words, the model inputs 
and PDF selections were `compiled in' and the executable had to be 
re-built every time those values changed.

The present command-line parser accepts so-called colon arguments 
(arguments starting with a `:') before the usual ones on the command 
line, as in:
\begin{alltt}
   run :\(\mathit{arg}\sb1\) :\(\mathit{arg}\sb2\) ... uuuuu 0,1000
\end{alltt}
The colon arguments are read into an array (sans colon) and handed to 
the model-initialization and luminosity-calculation subroutines:
\begin{verbatim}
   subroutine ModelDefaults(argc, argv)
\end{verbatim}
\begin{verbatim}
   subroutine LumiDefaults(argc, argv)
   integer argc
   character*128 argv(*)
\end{verbatim}
Note that, unlike in C (\Code{char **argv}), fixed-length strings are
passed in \Code{argv} since there are no pointers in Fortran 77.  It is 
up to the \Code{ModelDefaults} and \Code{LumiDefaults} subroutines to 
handle the arguments.  In Fortran it is furthermore no fatal error to
have no formal arguments in the \Code{ModelDefaults} and 
\Code{LumiDefaults} subroutines (as in previous FormCalc versions), so 
old code will compile and run without change.

\section{MSSM initialization via FeynHiggs}

The colon arguments of the previous section are immediately put to use 
for the initialization of the MSSM through FeynHiggs \cite{FeynHiggs}.
The default MSSM initialization is a stand-alone routine (\ie requires 
no external library to be linked), but is not quite as thorough as 
FeynHiggs when it comes to the corrections included \eg in the 
computation of the Higgs masses.

From FeynHiggs version 2.8.1 on not only the computational engine but 
the entire Frontend functionality is available through library routines 
so that the colon arguments can simply be passed to a FeynHiggs 
subroutine to make FeynHiggs initialize itself as if invoked from its 
own command-line Frontend.  The FormCalc-generated code inherits thus 
the ability to read parameter files in either native FeynHiggs or SLHA 
format, and of course obtains all MSSM parameters and Higgs observables 
from FeynHiggs. There is no duplication of initialization code this way, 
and moreover the parameters are consistent between the Higgs-mass and 
the cross- section calculations.

To use the FeynHiggs initialization, one chooses the 
model-initialization module \Code{model\_fh.F} instead of 
\Code{model\_mssm.F}.  The compiled code is invoked as
\begin{alltt}
   run :\(\mathit{parafile}\) \greyed{:\(\mathit{flags}\)} uuuuu 0,1000
\end{alltt}
The colon arguments are just the ones of the FeynHiggs Frontend: 
\textit{parafile} is the name of the parameter file and the optional 
\textit{flags} allows to override the default flags of FeynHiggs.

\section{Analytic tensor reduction}

Despite the hype that surrounds unitarity methods today, the 
Passarino--Veltman decomposition of tensor one-loop integrals 
\cite{PaVe} remains a valuable technique, also because it admits a fully 
analytic reduction.  The complete tensor reduction consists of two 
steps:
\begin{itemize}
\item The Lorentz-covariant decomposition of the tensors of the loop 
momentum appearing in the numerator into linear combinations of tensors 
constructed from $g_{\mu\nu}$ and the external momenta with coefficient 
functions, \eg
$$
\int\rd^4 q\frac{q_\mu q_\nu}{D_0 D_1} \sim
B_{\mu\nu} = g_{\mu\nu} B_{00} + p_\mu p_\nu B_{11}\,.
$$
This part has always been performed in FormCalc, as the actual tensors 
are rather unwieldy objects for further evaluation.

\item Solving the linear system that determines the coefficient 
functions, \ie expressing the coefficient functions through scalar 
integrals.
\end{itemize}

FormCalc has for long included the add-on \Code{FormCalc`btensor`} 
package which analytically reduces one- and two-point functions when 
loaded, but higher-point functions could be reduced only indirectly 
through FeynCalc \cite{FeynCalc}, \ie the user had to convert/save the 
amplitudes with \Code{FeynCalcPut}, run FeynCalc in a different 
Mathematica session, and load the reduced expressions into FormCalc 
again with \Code{FeynCalcGet}.  This procedure was not only suboptimal 
in terms of user-friendliness but also did not take advantage of the 
field levels of FeynArts, \ie FeynCalc always operated on the fully 
inserted amplitudes rather than the (typically much fewer) Generic 
amplitudes.

The analytic tensor reduction is meanwhile properly available in 
FormCalc and can be turned on through the option
\begin{alltt}
   CalcFeynAmp[..., PaVeReduce \(\to\) True]
\end{alltt}
Our code implements the reduction formulas of Denner and Dittmaier 
\cite{TensRed}.  While these are fully worked out, it nevertheless took 
considerable effort to program them in FORM due to at first sight 
trivial issues, \eg that there is no straightforward way to obtain the 
$N$-th argument of a function.  Adding the reduction code to the 
Mathematica part of FormCalc instead was not an option, however, as we 
wanted to operate on the Generic amplitudes, before the substitution of 
the insertions, and this happens in FORM.

Inverse Gram determinants, which appear as a by-product of inverting the 
coefficient-function system, may lead to instabilities in the numerical 
evaluation later on and therefore FormCalc tries to cancel them as much 
as possible.  The ones that cannot be cancelled immediately are returned 
as \Code{IGram[$x$]} ($= 1/x$) and so can easily be found and processed 
further in Mathematica.

\section{Auxiliary functions for operator matching}

As numerical calculations are done mostly using Weyl-spinor chains, 
there has been a paradigm shift from FormCalc 6 on for Dirac chains,
to make them better suited for analytical purposes, \eg the extraction 
of Wilson coefficients.

Several \Code{CalcFeynAmp} options allow to arrange Dirac chains in 
almost any prescribed order so that the coefficient multiplying a 
product of Dirac chains can be read off easily.
\begin{itemize}
\item The \Code{Antisymmetrize} option allows the choice of completely 
antisymmetrized Dirac chains, \ie \Code{DiracChain[-1,\,$\mu$,\,$\nu$] = 
$\frac 12[\gamma_\mu, \gamma_\nu]$}.

\item The \Code{FermionOrder} option implements Fierz methods for Dirac 
chains, allowing the user to force fermion chains into any desired 
order, \eg \Code{FermionOrder $\to$ \{2,1,\,4,3\}} produces fermion 
chains of the type $\bra{2}\cdots\ket{1} \bra{4}\cdots\ket{3}$.  
Alternately, \Code{FermionOrder $\to$ Colour} brings the spinors into 
the same order as the external colour indices.  If only simplification 
is sought, \Code{FermionOrder $\to$ Automatic} chooses a lexicographical
ordering.

\item The \Code{Evanescent} option introduces for every application of 
the Fierz identity a term of the form \Code{Evanescent[}\textit{original 
operator}, \textit{Fierzed operator}\Code{]} with the help of which one 
can detect problems due to the application of the Fierz identities.
\end{itemize}

\section{Built-in parallelization in Cuba}
\label{sect:cuba}

Cuba is a library for multidimensional numerical integration which is 
included in FormCalc but of course can be used independently, too.
Only the Mathematica interface was able to compute in parallel so far,
by redefining the function \Code{MapSample} with \eg \Code{ParallelMap}.
In the latest release, Cuba 3, we added parallelization also to the C/C++
and Fortran interfaces.

We attempt no parallelization across the network, say via MPI.  That is, 
we restrict ourselves to parallelization on one computer, using 
operating-system functions only, hence no extra software needs to be 
installed.  A common setup these days, even on laptops, is a single CPU 
with a number of cores, typically 4 or 8.  Utilizing many more compute 
nodes, as one could potentially do with MPI, is more of a theoretical 
option anyway since the speed-ups cannot be expected to grow linearly.

We use \Code{fork}/\Code{wait} rather than the \Code{pthread*} 
functions.  The latter are slightly more efficient at communicating data 
between parent and child because they share the same memory space, but 
for the same reason they also require a reentrant integrand function, 
and apart from the extra work this takes, a programmer may not even have 
control over reentrancy in his language, \eg Fortran's I/O is typically 
non-reentrant.  \Code{fork} on the other hand creates a completely 
independent copy of the running process and thus works for any integrand 
function with almost no restrictions (buffered I/O to a common file, for 
example, usually leads to unexpected results when executed 
concurrently).

Because a \Code{fork} is moderately expensive even on Linux with its 
efficient copy-on-write implementation, we use the `spinning threads' 
method, \ie a Cuba routine forks its workers once upon entry and 
afterwards starts and stops them by sending data or collecting results.

\begin{figure}[h]
\includegraphics[width=\hsize]{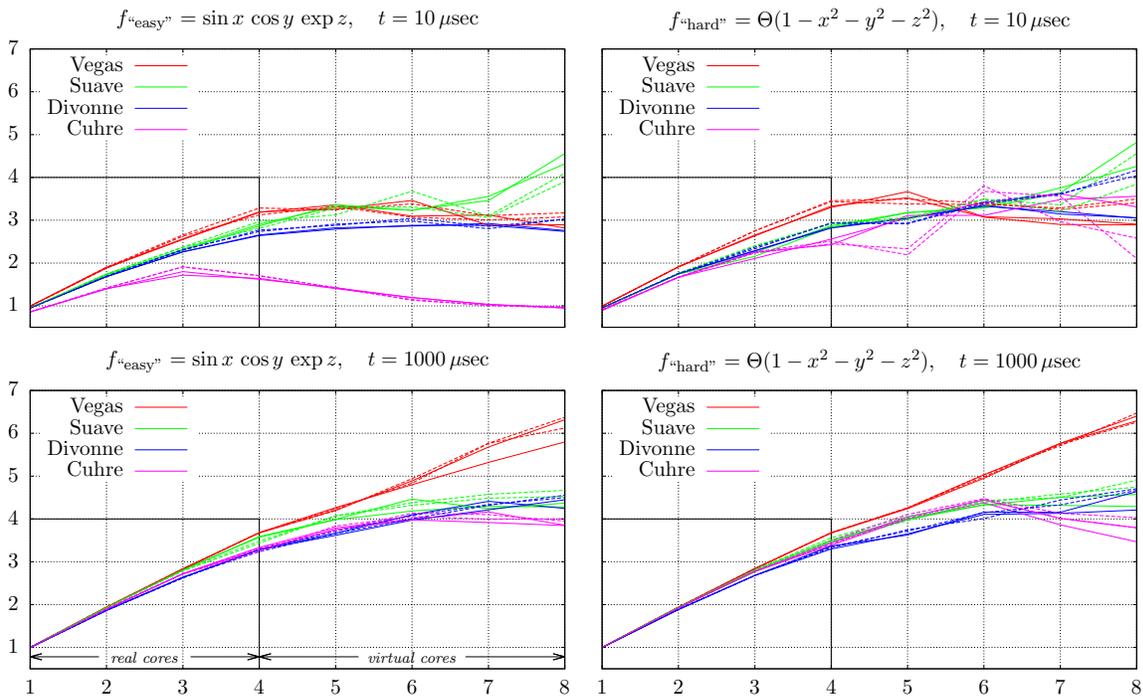}
\caption{\label{fig:speedups}Cuba speed-ups for a three-dimensional 
integral on an i7-2600 Linux system (3.1.10) with 4 real/8 virtual 
(hyperthreaded) cores.  The vertical line at 4 cores marks the 
cross-over.  The requested accuracy is $10^{-4}$ in all cases.
\textbf{Left column:} `easy' integrand.
\textbf{Right column:} `hard' integrand.
\textbf{Top row:} `fast' integrand ($10\,\mu$sec).
\textbf{Bottom row:} `slow' integrand ($1000\,\mu$sec per evaluation).
\textbf{Solid line:} shared memory,
\textbf{Dashed line:} socketpair communication
(two curves each to show fluctuations in timing measurements).
Note that also in the one-core case a parallel version is used (one 
master, one worker), which explains why the timings normalized to the 
serial version are below 1, in the top row visibly so.
What appears to be a drastic underperformance of Cuhre in the upper left 
panel can in fact be attributed to Cuhre's outstanding efficiency: it 
delivers a result correct to almost all digits with around 300 samples.  
In such a case, Cuba may for efficiency choose not to fill all available 
cores and relative to the full number of cores this shows up as a 
degradation.}
\end{figure}

The communication of samples to and from the workers happens through IPC 
shared memory (\Code{shmget} and friends), or if that is not available,
through a \Code{socketpair}.  Remarkably, the former's anticipated 
performance advantage turned out to be hardly perceptible.  Possibly 
there are cache coherence issues introduced by several workers writing 
simultaneously to the same shared-memory area.

Changing the number of cores to use does not require a re-compile, which 
is particularly useful as the program image should be able to run on 
several computers (with possibly different numbers of cores) 
simultaneously.  Cuba determines the number of cores from the 
environment variable \Code{CUBACORES}, or if this is unset, takes the 
idle cores on the present system (total cores minus load average).  That 
is, unless the user explicitly sets \Code{CUBACORES}, a program calling 
a Cuba routine will automatically parallelize on the available cores.  A 
master process orchestrates the parallelization but does not count 
towards the number of cores, \eg \Code{CUBACORES = 4} means four workers 
and one master.  Very importantly, the samples are generated by the 
master process only and distributed to the workers, such that random 
numbers are never used more than once.

Parallelization entails a certain overhead as usual, so the efficiency 
will depend on the `cost' of an integrand evaluation, \ie the more 
`expensive' (time-consuming) it is to sample the integrand, the better 
the speed-up will be.  To give an idea of the values that can be 
attained, Fig.\ \ref{fig:speedups} shows the speed-ups for an `easy' and 
a `hard' one of the 11 integrands of the demo program included in the 
Cuba package for two different integrand delays.  To tune the `cost' of 
the integrands, we introduced a calibrated delay loop into the integrand 
functions (which are simple one-liners and for our purposes `infinitely' 
fast).  The calibration and the timing measurements are rather delicate 
and shall not be discussed here.

The first, expected, observation is that parallelization is worthwhile 
only for not-too-fast integrands.  This is not a major showstopper, 
however, as many integrands in particle physics (one-loop 
cross-sections, for example) safely fall into the 
1000-$\mu$sec-and-beyond category.

The second observation is that parallelization works best for 
`simple-minded' integrators, \eg Vegas.  The `intelligent' algorithms 
are generally much harder to parallelize because they don't just do 
mechanical sampling but take into account intermediate results, make 
extra checks on the integrand (\eg try to find extrema), etc.  This is 
particularly true for Divonne, which was originally a recursive 
algorithm and thus hard to distribute.  It took significant effort to 
un-recurse the algorithm and lift the speed-up curve even this far above 
1, but still Divonne is lagging somewhat in parallelization efficiency.  
Then again, the `intelligent' algorithms are usually faster to start 
with (\ie converge with fewer points sampled), which compensates for the 
lack of parallelizability.

\section{Summary}

FormCalc 7.5 (\Code{http://feynarts.de/formcalc}) has many new and 
improved features, most notably OPP methods, C-code generation, the link 
with FeynHiggs, and analytic tensor reduction.  Cuba 3 
(\Code{http://feynarts.de/cuba}), included also in FormCalc, 
parallelizes integrations automatically and achieves decent speed-ups 
for typical cross-section integrands.

\medskip
\raggedright


\begin{thebibliography}{9}

\bibitem{FormCalc}
Hahn T, P\'erez-Victoria M,
1999, \textsl{Comput.\ Phys.\ Commun.} \textbf{118} 153
[hep-ph/9807565].
  %%CITATION = CPHCB,118,153;%%

\bibitem{FeynArts}
Hahn T,
2001, \textsl{Comput.\ Phys.\ Commun.} \textbf{140} 418
[hep-ph/0012260].
  %%CITATION = CPHCB,140,418;%%

\bibitem{Cuba}
Hahn T,
2005, \textsl{Comput.\ Phys.\ Commun.} \textbf{168} 78
[hep-ph/0404043].
  %%CITATION = HEP-PH/0404043;%%

\bibitem{OPP}
Ossola G, Papadopoulos C, Pittau R,
2007, \textsl{Nucl.\ Phys.\ B} \textbf{763} 147
[hep-ph/0609007].
  %%CITATION = NUPHA,B763,147;%%

\bibitem{CutTools}
Ossola G, Papadopoulos C, Pittau R,
2008, \textsl{JHEP} \textbf{0803} 042
[arXiv:0711.3596].
  %%CITATION = JHEPA,0803,042;%%

\bibitem{Samurai}
Mastrolia P, Ossola G, Reiter T, Tramontano F,
2010, \textsl{JHEP} \textbf{1008} 080
[arXiv:1006.0710].
  %%CITATION = JHEPA,1008,080;%%

\bibitem{PaVe}
Passarino G, Veltman M,
1979, \textsl{Nucl.\ Phys. B} \textbf{160} 151.
  %%CITATION = NUPHA,B160,151;%%

\bibitem{FCabbr}
Hahn T,
2010, \textsl{PoS ACAT 2010} 078
[arXiv:1006.2231]
  %%CITATION = POSCI,ACAT2010,078;%%

\bibitem{FCopt}
Hahn T,
2003, \textsl{Nucl.\ Phys.\ Proc.\ Suppl.} \textbf{116} 363
[hep-ph/0210220]
  %%CITATION = HEP-PH/0210220;%%

\bibitem{FeynHiggs}
Frank M, Hahn T, Heinemeyer S, Hollik W, Rzehak H, Weiglein G,
2007, \textsl{JHEP} \textbf{0702} 047
[hep-ph/0611326].
  %%CITATION = JHEPA,0702,047;%%

\bibitem{FeynCalc}
Mertig R, B\"ohm M, Denner A,
1991, \textsl{Comput.\ Phys.\ Commun.} \textbf{64} 345.
  %%CITATION = CPHCB,64,345;%%

\bibitem{TensRed}
Denner A, Dittmaier S,
2006, \textsl{Nucl.\ Phys.\ B} \textbf{734} 62
[hep-ph/0509141].
  %%CITATION = NUPHA,B734,62;%%

\end{thebibliography}
\end{document}